\documentclass[C4paper, 21pt]{article}
\usepackage{amsmath,amsthm,amssymb}
\usepackage{graphicx}
\usepackage{color}
\usepackage{dsfont}
\usepackage{epstopdf}
\usepackage[hang]{subfigure}
\usepackage{natbib}
\usepackage{hyperref}
\usepackage{booktabs}
\usepackage{textcomp}
\usepackage{bbm}
\usepackage{multirow}

\usepackage[top=2cm, bottom=2cm, left=2cm, right=2cm]{geometry}
\usepackage{algorithm}
\usepackage{algorithmicx}
\usepackage{algpseudocode}

\usepackage{ulem}
\usepackage{color}

\newtheorem{theorem}{Theorem}

\newtheorem{prop}{Proposition}

\definecolor{red}{RGB}{139,0,18}
\definecolor{lightred}{RGB}{186,25,31}
\definecolor{blue}{RGB}{0, 0, 255}
\definecolor{lightblue}{RGB}{69,100,139}
\renewcommand\emph[1]{{\color{red}\itshape #1}}

\newcounter{rmk}
\newcommand\rmk[1]{\vspace*{1mm} \par \stepcounter{rmk}{\noindent \bf Remark \thermk}. {#1}\vspace*{1mm}}

\newcommand\e{\epsilon}

\newcommand\mR{\mathds{R}}

\newcommand{\normm}[1]{{\vert\kern-0.25ex\vert\kern-0.25ex\vert #1
\vert\kern-0.25ex\vert\kern-0.25ex\vert}}

\title{A model-free feature selection technique of feature screening and random forest based recursive feature elimination}
\author{Siwei Xia\thanks{School of Science, Civil Aviation Flight University of China} ~~ and ~~ Yuehan Yang\thanks{School of Statistics and Mathematics, Central University of Finance and Economics}~\thanks{Corresponding Author. Email: yyh@cufe.edu.cn.}
}
\date{}

\begin{document}
\maketitle
\begin{abstract}
In this paper, we propose a model-free feature selection method for ultra-high dimensional data with mass features. This is a two phases procedure that we propose to use the fused Kolmogorov filter with the random forest based RFE to remove model limitations and reduce the computational complexity. The method is fully nonparametric and can work with various types of datasets. It has several appealing characteristics, i.e., accuracy, model-free, and computational efficiency, and can be widely used in practical problems, such as multiclass classification, nonparametric regression, and Poisson regression, among others. We show that the proposed method is selection consistent and $L_2$ consistent under weak regularity conditions. We further demonstrate the superior performance of the proposed method over other existing methods by simulations and real data examples.

\end{abstract}

\vspace*{4mm}

\section{Introduction}
Due to the development of data technology, feature selection plays an important role in both statistics and machine learning. High dimensional and ultra-high dimensional datasets are widely used in many fields, such as finance, image recognition, text classification, etc. Although more detailed information is provided with the increase of dimensions, the existence of a large number of redundant features weakens the generalization ability of models and increases the difficulty of data analysis \citep{jain2000statistical}.
Thus, the efficiency of feature selection is crucial, as it focuses on choosing a small subset of informative features that contain the information of data and determine the source of the specific concerns derived from the study. In many data analyses, feature selection is a significant and frequently used dimensionality reduction technique and is often considered a key preprocessing step in data analysis for model benefits, such as interpretability, accuracy, lower computational costs, and less prone to overfitting \citep{sheikhpour2017survey,khaire2022stability}.

Numerous studies on feature selection have been conducted. Generally speaking, feature selection can be divided into three categories: embedded, filter, and wrapper \citep{guyon2003introduction,guyon2008feature}.
Embedded approaches incorporate model learning by 1) using the objective function optimization, 2) calculating the change during the learning process, and 3) selecting the set of variables that has the best solution as the best model. This kind of approach selects the variables during the learning process. The two typical examples of this kind are the Lasso \citep{tibshirani1996lasso} with the $l_1$ regularization penalty and the Decision trees \citep{cho2011decision}. Penalized regularizations that shrink estimates by penalty functions, e.g., Lasso, SCAD \citep{fan2001variable}, Elastic Net \citep{zou2005elastic}, have been studied extensively for high dimensional data. These methods estimate and select features simultaneously and are computationally efficient. Many researchers have investigated the algorithms and statistical properties of these methods. Examples include the least angle regression \citep{efron2004lars}, coordinate descent algorithm \citep{friedman2007pathwise,friedman2010regularization}, theoretical guarantee for the Lasso \citep{zhao2006lasso}, etc. However, regularization methods are limited by model assumptions and are mostly applied to regression problems. Further, the estimation accuracy and computational cost are always influenced by the selection of tuning parameters.

Filter, also known as the variable ranking technique, calculates a particular statistical measure for each variable. They rank features and select the optimal subset according to the predetermined selection criteria.
These techniques are often used as pre-selection strategies that are independent of the latter applied machine learning algorithms \citep{janecek2008relationship}.
They are practically free as they do not rely on inductive algorithms.
Fisher score \citep{furey2000support} and Person correlation \citep{miyahara2000collaborative} are two classical ranking criteria. Two nonlinear approach filter techniques, Joint mutual information maximization and Normalized joint mutual information Maximization \citep{bennasar2015feature}, use mutual information and the maximum of the minimum criterion to produce the trade-off between accuracy and stability.
The characteristics of the filter technique are fast, simple, and efficient \citep{zhou2019feature,wang2021feature}.

Feature screening as a filter technique is an essential way to solve the problem caused by ultra-high dimensionality. It is less ambitious because it only aims to discover a majority of irrelevant variables. In other words, it finds a set that contains important features; in the meantime, the set also allows many irrelevant variables.

This concept is first illustrated in \cite{fan2008sure}, which proposed a feature screening method called sure independence screening (SIS).
This paper aimed to remove the redundant features by ranking their marginal Person correlations and provided the theoretical results called the sure independence screening property, showing that the remaining feature set contains all the important variables with high probability. The facility, effectiveness, and promising numerical performance of SIS make it popular among ultra-high dimensional analyses \citep{fan2020statistical,liu2020variable}. After that, feature screening has been applied to many problems, including parametric models, e.g., \cite{fan2010sure, zhao2012principled, xu2014ultrahigh}, and semiparametric or nonparametric models, e.g., \cite{fan2011nonparametric, cheng2014nonparametric, cheng2016forward, chu2016feature, chu2020feature}. The main drawback of this kind of filter technique is that the selection process does not take into consideration the performance of the learning model.
The previous studies mentioned above also have model limitations and cannot select the active set accurately.

The last category of feature selection is the wrapper, which searches for the optimal model in every possible combination of the available features by computing the model performance, like a search problem.
Each possible model is calculated based on the model accuracy, and we aim to choose the best model with the best model performance. Wrappers are also widely studied for their simplicity, availability, and generalization. The commonly used wrapper methods are forward selection based approaches \citep{blanchet2008forward, cheng2016forward}, backward selection based approaches \citep{pierna2009backward}, etc. However, these methods are often concerned by the computational cost, and they are not suitable for ultra-high dimensional data. To solve the problems, \cite{borboudakis2019forward} proposed a Forward-backward selection method with early dropping that significantly improves the running time. \cite{zheng2020building} proposed a sequential conditioning approach by dynamically updating the conditioning set with an iterative selection process under the framework of generalized linear models. \cite{honda2021forward} proposed forward variable selection procedures in ultra-high dimensional generalized varying coefficient models and established their theoretical properties. The above three approaches are suitable for ultra-high dimensionality but still model-based feature selection procedures.

Based on the existing results, we summarize that an appealing feature selection approach should satisfy the following three properties:
\begin{itemize}
\item \textbf{Accuracy}, which means that the subset consisting of informative features can be correctly selected. This is a basic requirement, and most methods have desirable accuracy under suitable conditions.
\item \textbf{Model-free}, i.e., it can be implemented without specifying a specific model. Specifying a model is challenging for empirical analysis. Recently, the model-free feature selection method becomes a hot research topic for its generalization and validity.
\item \textbf{Computational efficiency}, especially for ultra-high dimensional dataset that is usually time-consuming.
\end{itemize}

For the second property, model-free feature screening is first proposed by \cite{zhu2011model}. After that, \cite{he2013quantile} proposed a quantile-adaptive model-free feature screening framework for high dimensional heterogeneous data. \cite{mai2015fused} further developed the fused Kolmogorov filter for model-free feature screening with categorical, discrete, and continuous response. \cite{liu2022model} proposed a model-free and data-adaptive feature screening method, named PC-Screen, which is based on ranking the projection correlations between features and response variables. A state-the-art-of approach to wrapper methods without model restrictions is Recursive feature elimination (RFE), a sequential backward elimination, i.e., Support vector machine based Recursive feature elimination (SVM-RFE) \citep{guyon2002gene, lin2012support,guo2021feature},Random forest  based Recursive feature elimination (RF-RFE) \citep{gregorutti2017correlation, rumao2019exploration}, partial least squares based Recursive feature elimination (PLS-RFE) \citep{you2014pls}. Motivated by RFE, \cite{xia2022iterative} proposed an iterative model-free feature screening procedure named Forward recursive selection.

For the third property, the computational complexity of the filter is less than compared to the embedded and the wrapper techniques \citep{rumao2019exploration, seijo2017ensemble}. For the wrapper method, \cite{borboudakis2019forward} dramatically increased computational efficiency with early dropping. \cite{honda2021forward} and \cite{xia2022iterative} reduced computational consumption by adding a stopping rule that takes into account the model size. Unfortunately, none of the aforementioned feature selection approaches enjoys the three properties simultaneously.

Driven by the above problems, in this paper, we propose a model-free feature selection procedure for ultra-high dimensional datasets. This method proposes to use the fused Kolmogorov filter and the random forest based RFE, thus we abbreviate it as FK-RFE.
In the first filter phase, a ranking of features is returned by assigning a level of relevance to each feature, and the threshold value determines the retained features. In the second wrapper phase, successive subsets of features, generated according to a predefined search strategy, are evaluated according to an optimality criterion until a final subset of selected features is obtained. The contributions of this paper are as follows.

First, a model-free approach is developed for different types of ultra-high dimensional datasets. We propose to use the fused Kolmogorov filter to remove model assumptions and data assumptions. This method combines the advantages of the wrapper and filter strategies and is computationally efficient for mass features. We also show that the proposed method is appropriate for missing, outlier, and ultra-high dimensional data. The proposed method satisfies the appealing properties mentioned above: accuracy, model-free, and computational efficiency.

Second, the theoretical properties of model-free algorithms are always hard to obtain. To fill this gap, we establish the convergence theory of the proposed algorithm, proving that the procedure is selection consistent and $L_2$ consistent under mild conditions.

Third, we compare the proposed method with several existing methods in different models, such as the generalized linear model, the Poisson regression model, etc., under high and ultra-high dimensional settings. Both simulations and applications show the suitability and efficiency of the proposed approach.

The remainder of this paper is organized as follows. Section 2 describes the proposed method, the algorithm, and its advantages. Section 3 illustrates the theoretical properties. Sections 4 and 5 present the simulation and application results. Section 6 concludes the paper.

\section{Method}
\begin{figure}[htbp]
\centering
\includegraphics[width=1\textwidth,height=\columnwidth]{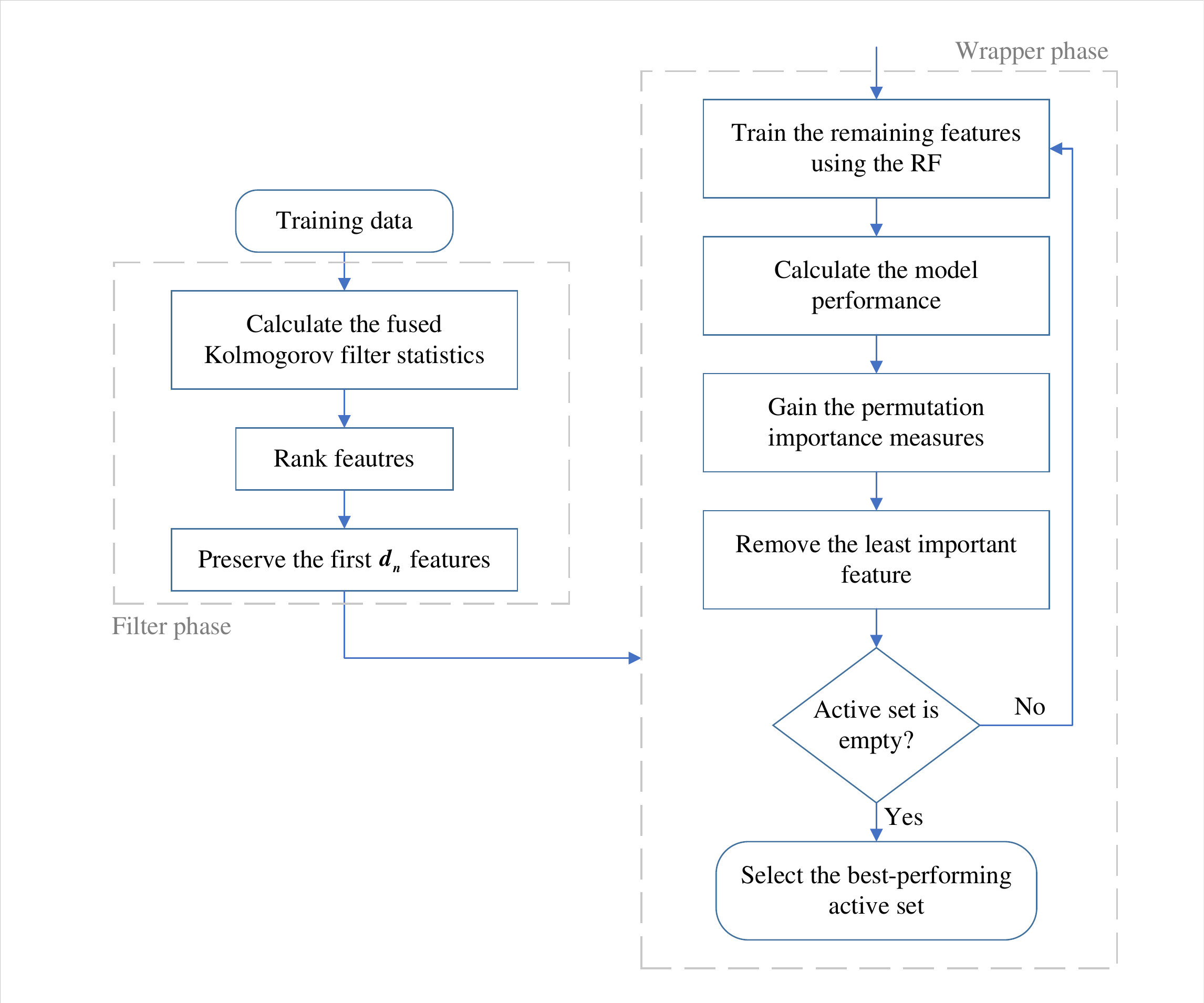}
\caption{The flowchart of the FK-RFE}
\label{figflowchart}
\end{figure}

In this section, we introduce the proposed mode-free feature selection procedure, FK-RFE. This method incorporates a filter phase and wrapper phase possessing the advantages of the feature screening, recursive feature elimination, and random forest. In the following we show that this technique is efficient and can be applied to various data. For simplicity of description, we first consider a supervised problem with a response $Y$, predictors $X=(X_1,\dots, X_p)$ and the following model framework:
\begin{align}\label{model1}
Y=f(X)+\e,
\end{align}
where $f$ is a measurable function and can be any model, e.g., parametric, semiparametric, or nonparametric model. $\e$, a noise term, is independent of predictor $X_j$ with $E(\e)=0$ and $Var(\e)=\sigma^2\in (0,\infty)$. When the dimension $p$ becomes very large, a reasonable requirement is the sparsity assumption that only a small subset of variables is responsible for modeling $Y$. The pseudo-code of FK-RFE with the execution process is given in the following Algorithm 1, and the flowchart is given in Figure~\ref{figflowchart}.

\begin{algorithm}\label{algorithm}
\caption{FK-RFE}
\hspace*{0.02in} {\bf Filter phase:}
\begin{algorithmic}[1]
    \State Rank the features according to the fused Kolmogorov filter statistic $\hat K_j$, $j=1,\dots,p$.
    \State Obtain the reduced set
    \[ V_0=\{j: |\hat K_j| \text{ is among the first } d_n \text{ largest of all}\}.\]
\end{algorithmic}
\hspace*{0.02in} {\bf Wrapper phase:}
\begin{algorithmic}[1]
    \For{all the remaining features}
    \State Train the model using the random forest.
    \State Calculate the model performance.
    \State Calculate the permutation importance measures.
    \State Update $V_{l-1}$ to $V_l$ by eliminating the least important feature.
    \State Update $l=l+1$.
    \State Continue until no features left.
    \EndFor
    \State Choose the active set $V_l$ with the best model performance as the best set.
\end{algorithmic}
\end{algorithm}

The proposed algorithm consists of a filter phase and a wrapper phase. In this algorithm, we first use a feature screening technique, the fused Kolmogorov filter, to remove numerous uninformative features and obtain a reduced active set $V_0$ which contains the true model. Then, we use the random forest to train the model and rank features by the permutation importance measure. During this step, we update the active set by eliminating the least significant feature. In each iteration, we rerank the remaining features by recalculating the permutation importance measure, for it works better than the one without reranking \citep{svetnik2004application,diaz2006gene}. The optimal subset is determined based on the best model performance. The details and advantages of this method are discussed in the following.

\subsection{The first phase: filter}
In this part, we introduce the first screening phase of the proposed method, the fused Kolmogorov filter, which is first introduced by \citet{mai2015fused} for model-free feature screening.
It is shown to enjoy the sure screening property under weak regularity conditions and is a powerful technique when covariates are strongly dependent on each other.  We propose using it to deal with a variety of datasets such as parametric and nonparametric regression. In the proposed algorithm, FK-RFE calculates the fused Kolmogorov filter statistics for all of the features and selects first $d_n$ features according to the values, where $d_n$ denotes the number of features and often relates to the sample size. The fused Kolmogorov filter screening set is defined as
\begin{align*}
V_0=\{1\leqslant j \leqslant p: \hat K_j \text{ is among the first } d_n \text{ largest of all}\},
\end{align*}
where $\hat K_j$ is the fused Kolmogorov filter statistic among $j$th predictor $X_j$.

We elaborate on the definition of $\hat K_j$ below. Suppose that we have $N$ different partitions of the value of the response,  $G_t$ ($t = 1,\dots, N$) is the $t$th partition with $g_t$ slices. If $Y$ is continuous, as \cite{mai2015fused} suggested, we set
\begin{align*}
G_t=\{ [a_{l-1},a_l) : a_{l-1}< a_l \text{ for } l=1,\dots,g_t, \text{and} \cup_{l=1}^{g_t} [a_{l-1},a_l)=\mR\},
\end{align*}
where $a_0=-\infty$, $a_{g_t}=+\infty$ and the interval $[a_0, a_1)=(-\infty, a_1)$.
The partition $G_t$ should contain the intervals bounded by the $\frac{1}{g_t}$th sample quantiles of $Y$.
For a given partition $G_t$, $K^{G_t}_j$ denotes the maximum value of the difference between conditional distributions of $X_j$ given different values of $Y$, defined as
\begin{align*}
K_j^{G_t}=\max_{l,\gamma} \sup_{x} |F_j(x|H_j=l)-F_j(x|H_j=\gamma)|,
\end{align*}
where $H_j$ is random variable taking values on $\{1,\dots,g_t\}$, $H_j=l$ if $y_i$ is in the $l$th slice, and $F_j$ denotes the generic cumulative distribution function (CDF) for $X_j$, $F_j(x|H_j=l)=P(X_j \leqslant x|H_j=l)$, $l=1,\dots,g_t$.
The more important $X_j$ is, the larger this difference is.
Naturally, the empirical version of $K_j^{G_t}$ based on a sample $(x_{ij},y_i)$, $i=1,\dots,n$, is defined as
\begin{align*}
\hat K_j^{G_t}=\max_{l,\gamma} \sup_{x} |\hat F_j(x|H_j=l)-\hat F_j(x|H_j=\gamma)|,
\end{align*}
where $\hat F_j(x|H_j=l)=(1/n_l)\sum_{\{H_j=l\}} 1(x_{ij}\leqslant x)$ with $n_l$ denotes the sample size of $\{H_j=l\}$. Then, the fused Kolmogorov filter statistic is the sum of $N$ different partitions which integrates several slicing schemes and is defined as
$$\hat K_j=\sum_{t=1}^{N}\hat K_j^{G_t}.$$
The main idea of this filter is that $X$ and $Y$ are independent if and only if the conditional distributions of $X$ given different values of $Y$ remain equal, and it performs better than others who rely on a single slicing scheme. In practice, $g_t \leqslant \lceil \log n \rceil$ for all $t$ is chosen so that each slice contains a sufficient sample size for all slicing strategies. Furthermore, if $Y$ is a multi-level categorical variable, as $Y=1,\dots,g$, then we only use one partition, and the partition is simply done according to $Y$'s level. That is, $G=\{1,\dots,g\}$. In this case, we set $H=Y$, i.e., $H=l$ if $Y=l$.

$d_n$ is a common parameter for the filter technique to control the model size. Similarly,
to contain the true model, we tend to use a reasonably large $d_n$. For ultra-high dimensional sparse models, we often assume that the number of important features is less than the sample size $n$. A conservative choice can be $d_n=a \lceil n/\log n \rceil$, where $a$ is some constant \citep{mai2015fused,fan2008sure}.

We use the fused Kolmogorov filter as the first screening phase,
which has four main advantages: 1) Free model restrictions, allowing the proposed method to be widely used for various types of data. 2) Fast and simple, especially for ultra-high dimensional data. 3) Theoretical guarantee. In the next section we show that the obtained subset from FKF-RFE contains all the relevant variables. 4) Screening efficiency, i.e., after screening, the model size is controlled by $d_n$.

\subsection{The second phase: wrapper}
In this part, we introduce the second half of the proposed method. In this phase, we use a backward strategy called Recursive feature elimination (RFE). At each step of this strategy, the variable importance ranking is updated under the current model, and the feature with the lowest importance measure is removed from the active set. This strategy is first introduced by \citet{guyon2002gene} for SVM and has gained popularity in many fields, i.e., gene selection \citep{duan2005multiple,mundra2009svm}, medical diagnosis \citep{richhariya2020diagnosis,senan2021diagnosis}. In the proposed algorithm, we obtain a reduced set from the first filter phase and use this reduced set as the initial set of the RFE in the second wrapper phase. In this phase, we use the random forest to train the model and obtain the permutation importance measure for variable ranking. MSE (for continuous response) or out-of-bag error (for multi-level categorical response) is used as the model selection criteria. Now we briefly introduce the techniques used in this part.

Random forest is an ensemble learning approach and operates on the mechanism of the bagging method \citep{breiman2001random}. It consists of a collection of decision trees. Let $D=\{(x_1,y_1),\dots, (x_n,y_n)\}$ be the sample set of $(X,Y)$.  $\hat f$ is an estimate of $f$ used to predict $Y$. The trees are constructed using $M$ bootstrap samples $D_1,\dots, D_M$ of $D$. The learning rule of random forest is the aggregation of all the tree-based estimators denoted by $\hat f_1,\dots, \hat f_M$ where the aggregation is calculated based on the average of the predictions $\hat f=\frac{1}{M}\sum_{m=1}^{M} \hat f_m$.

Random forest access the relevance of a predictor by using variable importance measure. The permutation importance measure \citep{breiman2001random,gregorutti2017correlation} is adopted in this paper. This measure is based on the idea that a variable $X_j$ is relevant to $Y$ if the prediction error increases when we break the link between $X_j$ and $Y$, and this link can be broken by random permuting the observations of $X_j$. That is, for $j=1,\dots,p$, set $X_{(j)}=(X_1,\dots,X'_j,\dots,X_p)$ be the random vector in which $X'_{j}$ is an independent replication of $X_j$. The permutation importance measure is given by
\begin{align*}
I(X_j)=E[(Y-f(X_{(j)}))^2]-E[(Y-f(X))^2].
\end{align*}
Note that the random permutation also breaks the link between $X_j$ and other predictors. In other words, $X'_j$ is independent of $Y$ and other predictors $X_{j'}$, $j'\neq j$, simultaneously. Denote $\bar D_m=D \backslash D_m$ as the out-of-bag samples of $D_m$ to contain the observations which are not selected in $D_m$. Let $\bar D_m^j$ be the permuted out-of-bag samples by random permutations of the observations of $X_j$. The empirical permutation importance measure is expressed as
\begin{align}\label{reg empimportance}
\hat I(X_j)=\frac{1}{M}\sum_{m=1}^M[R(\hat f_m,\bar D^j_m)-R(\hat f_m,\bar D_m)],
\end{align}
where $R(\hat f_m,T)=(1/|T|)\sum_{i:(x_i,y_i)\in T} (y_i-\hat f_m(x_i))^2$ for sample set $T=\bar D^j_m$ or $T=\bar D_m$. The permutation importance measure is recalculated to rank the predictors in each iteration. In addition to other criteria, the permutation importance measure has proven to be effective for leading variable selection methods \citep{gregorutti2017correlation,ramosaj2019asymptotic}.

The advantages of using the random forest based RFE and the permutation importance measure in the second feature selection phase are two-fold: 1) This combination allows the proposed method to handle various types of data; 2) The proposed method achieves high accuracy based on the theoretical guarantee and numerical experience.

\subsection{Comparison with other methods}
In this part, we focus on discussing the characteristics of the proposed method and comparing it with other existing methods. The first characteristic of the FK-RFE is that this method requires neither model assumption nor data assumption, thus is efficient for various data and is suitable for nonparametric, semiparametric, and parametric problems. Additionally, using the random forest for training, the proposed procedure works well with noise and is also appropriate for missing, outlier, and ultra-high dimensional data. We also demonstrate these characteristics in the numerical experiments following. To acquire a better understanding of the FK-RFE, we explore the following distinctions between it and other methods:
\begin{itemize}
\item FK-RFE combines the advantages of both the wrapper and filter techniques; in the meantime, it avoids the disadvantages of the wrapper and filter according to the algorithm. For example, the first phase of the FK-RFE applies the high efficiency of dimension reduction from the filter; thus, in the second phase, we avoid the computational problem caused by the wrapper. In the meantime, the strategies from the wrapper help us highly improve the selection accuracy.

\item Compared to the model-based techniques, such as regularization approaches \citep{tibshirani1997lasso,fan2001variable,yang2016penalized}, model-based forward selections \citep{wang2009forward,ing2011stepwise,zheng2020building}, the proposed method is a model-free approach. It is suitable for a wider variety of data formats and requires fewer assumptions. Further, during the whole algorithm, we only need to pick one parameter $d_n$, which is not decisive for the method and also easy to calculate without cross-validation, BIC, or other parameter selection techniques.

\item Compared to other similar iterative algorithms, FK-RFE also has some advantages. As a comparison, RFE \citep{gregorutti2017correlation} is computationally consuming because the iteration starts from the whole variables until no variable remains. FRS \citep{xia2022iterative} is suitable for high dimensional data, but the number of iterations is determined by the number of samples, and the computational cost would be large when dealing with large data.
\end{itemize}

\section{Consistency analysis}
Noted that we consider sparse learning problems with the model framework \eqref{model1}, i.e.,
\[Y = f(X) +\e.\]
Considers a subset $S=\{j:\beta\neq 0\}$ with cardinality $|S|=q$ that much less than the dimension $p$. Formally speaking, we refer to a predictor $j\in S$ as informative or relevant. If a predictor $j$ belongs to the complement of $S$ ($j\in S^c=\{1,\dots,p\}\backslash S$), it is regarded as uninformative or unimportant.
In practice, a forest can only be created with a finite number of trees. On the other hand, in the theoretical analysis, it is generally assumed that $M$ tends to infinity. This is because when $M = \infty$, the predictors do not depend on the realization of the specific tree in the forest. To simplify the proof, we follow this assumption and consider the consistent property among the infinite forest. The infinite forest estimate is defined by $\bar f=E(\hat f)$.
By the law of large numbers,
\begin{align*}
\bar f=\lim_{M\to\infty}\hat f=\lim_{M\to\infty} \frac{1}{M}\sum_{m=1}^{M} \hat f_m,
\end{align*}
where more details can be found in \cite{breiman2001random,scornet2015consistency}. We consider the following regularity conditions, under which we can guarantee the convergence of Algorithm 1, that the FK-RFE is selection consistent and $L_2$ consistent.

\begin{description}
\item[C1.] There exists a set $S_1$ such that $S\subset S_1 $ and
\[\bigtriangleup_{S_1}=\min_{t}\big( \min_{j\in {S_1}} K_j^{G^o_t} - \max_{j\notin {S_1}} K_j^{G^o_t}\big)>0, \]
where $ K_j^{G^o_t}$ is $ K_j^{G_t}$ under oracle partition $G_t$ which contains the intervals bounded by the $1/g_t$th theoretical quantiles of $Y$.
\item[C2.]  For any $b_1$, $b_2$ such that $P(Y\in [b_1,b_2))\leqslant 2/min\{g_t\}$, we have
\[|F_j(x|y_1)-F_j(x|y_2)|\leqslant \frac{\bigtriangleup_{S_1}}{8},\]
for all $x$, $j$ and $y_1, y_2 \in [b_1,b_2)$.
\end{description}
\begin{theorem}\label{thm 1}
Suppose conditions C1-C2 hold. Assume the importance measure $\hat I(X_j)$ is an unbiased estimator of $I(X_j)$, i.e., $\lim_{M\to \infty}E(\hat I(X_j))=I(X_j)$, as $n\to \infty$, and the infinite random forest is $L_2$ consistent. Then the FK-RFE is selection consistent. That is, denoting $\hat S$ to be the set selected by the FK-RFE, we have
\begin{align*}
P(\hat S = S)\to 1, \text{as } n\to \infty.
\end{align*}
\end{theorem}

\rmk Conditions C1 and C2 follow the conditions C1 and C2 in \cite{mai2015fused}, which are used to guarantee the sure screening property of the fused Kolmogorov filter. Condition C1 guarantees that the predictors belonging to $S$ should be marginally important, which is also a regular assumption in existing marginal screening approaches. Condition C2 guarantees that the sample quantiles of $Y$ are close enough to the population quantiles of $Y$. Both conditions are very mild.

\rmk The requirement of importance measure is formally proven to be valid under some general assumptions in \cite{ramosaj2019asymptotic}. This ensures that the permutation importance measure of the informative predictor converges to a non-zero constant and that one of the uninformative predictors converges to 0 with probability. Thus, uninformative predictors are eliminated before informative ones. The permutation importance measure is widely studied in many researches, i.e., \cite{altmann2010permutation,gregorutti2017correlation,ramosaj2019asymptotic,xia2022iterative}.
For example, assuming an additive regression model, i.e., $f(X)=\sum_{j=1}^p f_j(X_j)$, \cite{gregorutti2017correlation} proposed that $I(X_j)=2Var(f_j(X_j))$. \cite{ramosaj2019asymptotic}  proved that, under more general assumptions and model \eqref{model1}, $I(X_j)$ equals $E[(f(X)-f(X_{(j)}))^2]$ for $j\in S$, or equals 0 for $j\in S^c$.

\rmk Another important requirement of Theorem \ref{thm 1} is the $L_2$ consistency requirement of the random forest estimator. Many references are dedicated to this result. For example, \cite{breiman2001random} proved an upper bound on the generalization error of forests based on the correlation and strength of the individual trees. \cite{denil2013consistency} proved the consistency of online random forest. \cite{scornet2015consistency} showed that the random forest is $L_2$ consistent in an additive regression framework. \cite{athey2019generalized} proposed generalized random forest and developed asymptotic and consistency theory. In this case, we omit the detailed proof of this requirement and refer readers to the above references.

\begin{prop}
Assume the infinite random forest is $L_2$ consistent. The FK-RFE is $L_2$ consistent too.
\end{prop}

\rmk Since the FK-RFE uses the random forest for prediction, the $L_2$ consistency of the FK-RFE is obtained directly from the $L_2$ consistency of the random forest estimator. We just state a proportion here without proof.

\section{Simulations}
In this section, we compare the FK-RFE with other methods on simulated datasets ranging from low latitude to ultra-high dimensional. In all the examples, we set sample size $n=100$ and the number of features $p=\{100,300,500,2000\}$. As \cite{mai2015fused} suggested, we consider $g_t=3,4$ in the fused Kolmogorov filter of the FK-RFE for $\lceil \log n \rceil=4$ and the threshold $d_n=\lceil \frac{n}{\log n} \rceil$. We consider four other feature selection methods for comparison, i.e., Recursive feature elimination (RFE) \citep{gregorutti2017correlation}, Forward recursive selection (FRS) \citep{xia2022iterative}, Lasso \citep{tibshirani1996lasso}, and Forward-backward selection withe early dropping (FBED) \citep{borboudakis2019forward}. We denote $X \sim \Sigma=(\sigma_{ij})_{p\times p}$, $\sigma_{ij}=0.5^{|i-j|}$, $\epsilon \sim N(0,1)$, and consider the following five models in this simulation study.
\begin{description}
\item Example 1. $Y=\exp (X\beta)+\epsilon$ with $\beta=(1,1,1,1,1,0,\dots,0)$.
\item Example 2. $Y^{1/9}=2.8X_1-2.8X_2+\epsilon$.
\item Example 3. $Y=(X_1+X_2+1)^3+\epsilon$.
\item Example 4. $Y=2(X_1+X_2)+2\tan (\pi X_3/2)+5X_4+\epsilon$.
\item Example 5.  $Y\sim \text{Poisson}(u)$, where $u=\exp (0.8X_1-0.8X_2)$, $X_j \sim t_2$ independently.
\end{description}

In different examples, the response $Y$ is generated by different models. Three different generalized linear models are used in Examples 1-3. Example 4 considers an additive model, and Example 5 considers a Poisson regression model used in \cite{mai2015fused}. Meanwhile, the number of relevant features in different models varies from 2 to 5. We use the True positive rate (TPR), True negative rate (TNR), Balanced accuracy, and the model size (number of selected features) to evaluate the selection performance of these methods. The results of Examples 1-5 are summarized in Tables \ref{tableE1}-\ref{tableE5}, respectively. Due to the limited computing power and the limitation of R software, the entries of RFE among $p=2000$ are missing. The values in parentheses indicate the sample standard deviation.

As shown in Tables \ref{tableE1}-\ref{tableE5}, the Balanced accuracy of FK-RFE performs well in all the examples, and the obtained model size is always smaller than most other methods while achieving high TPR and TNR. Specifically, FK-RFE selects 2-5 more variables in the model compared to the FBED, but the TPR is much better, indicating that the latter cannot choose the correct variables while FK-RFE can. Compared with the other three methods, i.e., the RFE, FRS, and Lasso, the Balanced accuracy and model size of FK-RFE are much better than the other three in all examples, while the model size is always less than a fifth. As shown in Table \ref{tableE5}, the Balanced accuracy, TPR, and TNR of FK-RFE are consistently better than other methods.
In Examples 1-4, some TPRs of the proposed method are lower with a difference of about 0.15; in the meantime, the TNRs always perform the best. This is because other methods tend to choose a rather large model, and easier to contain the relevant variables but with the cost of containing several irrelevant features. The FK-RFE, on the other hand, strikes a good balance between them. As the dimension increases and the number of irrelevant features increases, the selection accuracy of the proposed method is not affected. In summary, under various models and dimensions, FK-RFE always achieves high selection accuracy.

\begin{table}[!htp]
\centering
\small
\setlength\tabcolsep{21pt}
\caption{Performance comparison under Example 1.}
\label{tableE1}
\scalebox{0.83}{
\begin{tabular}{llcccccccc}
\hline\hline
Method  &   $p$ &   Balanced Accuracy       &   Model Size      &   TPR     &   TNR     \\\hline
\multirow{4}*{FK-RFE}   &   100 &   0.848   (0.12)  &   9.50    (5.78)  &   0.757   (0.26)  &   0.940   (0.05)  \\
&   300 &   0.868   (0.12)  &   9.36    (5.66)  &   0.755   (0.25)  &   0.981   (0.02)  \\
&   500 &   0.855   (0.12)  &   9.02    (5.52)  &   0.721   (0.25)  &   0.989   (0.01)  \\
&   2000&   0.867   (0.13)  &   9.89    (5.73)  &   0.737   (0.26)  &   0.997   (0.01)  \\\hline
\multirow{4}*{RFE}  &   100 &   0.696   (0.15)  &   48.24   (36.63) &   0.854   (0.23)  &   0.537   (0.38)  \\
&   300 &   0.660   (0.13)  &   186.98 (94.46) &    0.938   (0.14)  &   0.382   (0.32)  \\
&   500 &   0.689   (0.13)  &   286.24 (141.96) &   0.947   (0.13)  &   0.431   (0.29)  \\
&   2000&   --      &   --      &   --      &   --      \\\hline
\multirow{4}*{FRS}  &   100 &   0.699   (0.14)  &   50.04   (35.42) &   0.878   (0.22)  &   0.519   (0.37)  \\
&   300 &   0.873   (0.06)  &   62.66   (28.54) &   0.942   (0.15)  &   0.804   (0.10)  \\
&   500 &   0.903   (0.07)  &   66.70   (28.40) &   0.932   (0.16)  &   0.875   (0.06)  \\
&   2000&   0.922   (0.07)  &   72.57   (25.13) &   0.878   (0.15)  &   0.966   (0.01)  \\\hline
\multirow{4}*{Lasso}    &   100 &   0.613   (0.10)  &   62.18   (27.01) &   0.836   (0.17)  &   0.789   (0.28)  \\
&   300 &   0.723   (0.10)  &   79.72   (24.15) &   0.704   (0.19)  &   0.742   (0.08)  \\
&   500 &   0.730   (0.11)  &   57.88   (34.20) &   0.572   (0.24)  &   0.889   (0.07)  \\
&   2000&   0.725   (0.11)  &   87.52   (18.34) &   0.492   (0.22)  &   0.957   (0.01)  \\\hline
\multirow{4}*{FBED} &   100 &   0.642   (0.07)  &   3.40    (1.02)  &   0.304   (0.13)  &   0.980   (0.01)  \\
&   300 &   0.637   (0.06)  &   5.59    (1.25)  &   0.288   (0.12)  &   0.986   (0.01)  \\
&   500 &   0.621   (0.06)  &   6.99    (1.30)  &   0.254   (0.12)  &   0.988   (0.01)  \\
&   2000&   0.591   (0.06)  &   11.40   (1.29)  &   0.187   (0.12)  &   0.995   (0.01)  \\\hline

\end{tabular}}
\end{table}

\begin{table}[!htp]
\centering
\small
\setlength\tabcolsep{21pt}
\caption{Performance comparison under Example 2.}
\label{tableE2}
\scalebox{0.83}{
\begin{tabular}{llcccccccc}
\hline\hline

Method  &   $p$ &   Balanced Accuracy       &   Model Size      &   TPR     &   TNR     \\\hline
\multirow{4}*{FK-RFE}   &   100 &   0.805   (0.19)  &   7.31    (5.54)  &   0.670   (0.38)  &   0.939   (0.06)  \\
&   300 &   0.816   (0.18)  &   7.51    (5.39)  &   0.653   (0.37)  &   0.979   (0.02)  \\
&   500 &   0.755   (0.19)  &   7.30    (5.55)  &   0.523   (0.38)  &   0.987   (0.01)  \\
&   2000&   0.712   (0.17)  &   7.36    (5.40)  &   0.428   (0.35)  &   0.997   (0.01)  \\\hline
\multirow{4}*{RFE}      &   100 &   0.658   (0.18)  &   42.81   (38.48)     &   0.738   (0.36)  &   0.578   (0.39)  \\
&   300 &   0.567   (0.09)  &   251.34  (58.62)     &   0.970   (0.15)  &   0.163   (0.20)  \\
&   500 &   0.589   (0.11)  &   377.86  (106.34)    &   0.933   (0.20)  &   0.245   (0.21)  \\
&   2000&   --      &   --      &   --      &   --  \\\hline
\multirow{4}*{FRS}      &   100 &   0.630   (0.16)  &   49.49   (41.02)     &   0.750   (0.37)  &   0.510   (0.41)  \\
&   300 &   0.767   (0.14)  &   62.98   (33.86)     &   0.740   (0.34)  &   0.794   (0.11)  \\
&   500 &   0.771   (0.16)  &   71.76   (29.40)     &   0.683   (0.34)  &   0.859   (0.06)  \\
&   2000&   0.701   (0.19)  &   78.05   (25.96)     &   0.440   (0.38)  &   0.961   (0.01)  \\\hline
\multirow{4}*{Lasso}    &   100 &   0.679   (0.14)  &   62.62   (27.60)     &   0.978   (0.12)  &   0.781   (0.28)  \\
&   300 &   0.844   (0.11)  &   68.91   (27.82)     &   0.913   (0.20)  &   0.775   (0.09)  \\
&   500 &   0.850   (0.11)  &   88.44 (17.34)   &   0.875   (0.22)  &   0.826   (0.03)  \\
&   2000&   0.750   (0.17)  &   95.29   (8.60)  &   0.548   (0.33)  &   0.953   (0.01)  \\\hline
\multirow{4}*{FBED}     &   100 &   0.747   (0.15)  &   3.11    (1.07)  &   0.515   (0.30)  &   0.979   (0.01)  \\
&   300 &   0.675   (0.15)  &   5.47    (1.21)  &   0.365   (0.29)  &   0.984   (0.01)  \\
&   500 &   0.683   (0.16)  &   6.83    (1.24)  &   0.378   (0.33)  &   0.988   (0.01)  \\
&   2000&   0.621   (0.15)  &   11.55   (1.21)  &   0.248   (0.30)  &   0.994   (0.01)  \\\hline
\end{tabular}}
\end{table}

\begin{table}[!htp]
\centering
\small
\setlength\tabcolsep{21pt}
\caption{Performance comparison under Example 3.}
\label{tableE3}
\scalebox{0.83}{
\begin{tabular}{llcccccccc}
\hline\hline
Method  &   $p$ &   Balanced Accuracy       &   Model Size      &   TPR     &   TNR     \\\hline
\multirow{4}*{FK-RFE}   &   100 &   0.815   (0.13)  &   7.72    (5.40)  &   0.695   (0.28)  &   0.935   (0.05)  \\
&   300 &   0.818   (0.13)  &   7.64    (5.39)  &   0.658   (0.28)  &   0.979   (0.02)  \\
&   500 &   0.826   (0.14)  &   7.60    (5.38)  &   0.665   (0.29)  &   0.987   (0.01)  \\
&   2000&   0.822   (0.15)  &   7.13    (5.57)  &   0.648   (0.30)  &   0.997   (0.01)  \\\hline
\multirow{4}*{RFE}      &   100 &   0.710   (0.16)  &   42.01   (36.95)     &   0.833   (0.25)  &   0.588   (0.37)  \\
&   300 &   0.615   (0.14)  &   208.80  (94.32)     &   0.925   (0.18)  &   0.306   (0.32)  \\
&   500 &   0.591   (0.15)  &   360.16  (114.44)    &   0.903   (0.20)  &   0.280   (0.23)  \\
&   2000&   --      &   --      &   --      &   --  \\\hline
\multirow{4}*{FRS}      &   100 &   0.812   (0.19)  &   26.19   (35.75)     &   0.873   (0.22)  &   0.751   (0.36)  \\
&   300 &   0.869   (0.12)  &   40.88   (38.53)     &   0.870   (0.23)  &   0.869   (0.13)  \\
&   500 &   0.894   (0.11)  &   51.78   (36.58)     &   0.888   (0.23)  &   0.900   (0.07)  \\
&   2000&   0.923   (0.11)  &   46.56   (39.69)     &   0.868   (0.23)  &   0.978   (0.02)  \\\hline
\multirow{4}*{Lasso}    &   100 &   0.656   (0.17)  &   42.46   (41.46)     &   0.730   (0.34)  &   0.582   (0.42)  \\
&   300 &   0.694   (0.15)  &   30.80   (38.71)     &   0.488   (0.34)  &   0.900   (0.13)  \\
&   500 &   0.643   (0.15)  &   15.05   (30.59)     &   0.315   (0.33)  &   0.971   (0.06)  \\
&   2000&   0.801   (0.16)  &   40.97   (38.95)     &   0.623   (0.32)  &   0.980   (0.02)  \\\hline
\multirow{4}*{FBED}     &   100 &   0.814   (0.11)  &   2.44    (1.17)  &   0.640   (0.23)  &   0.988   (0.01)  \\
&   300 &   0.791   (0.11)  &   3.41    (1.54)  &   0.590   (0.21)  &   0.993   (0.01)  \\
&   500 &   0.775   (0.11)  &   3.85    (1.56)  &   0.555   (0.23)  &   0.995   (0.01)  \\
&   2000&   0.753   (0.12)  &   4.84    (1.51)  &   0.508   (0.23)  &   0.998   (0.01)  \\\hline

\end{tabular}}
\end{table}

\begin{table}[!htp]
\centering
\small
\setlength\tabcolsep{21pt}
\caption{Performance comparison under Example 4.}
\label{tableE4}
\scalebox{0.83}{
\begin{tabular}{llcccccccc}
\hline\hline
Method  &   $p$ &   Balanced Accuracy       &   Model Size      &   TPR     &   TNR     \\\hline
\multirow{4}*{FK-RFE}   &   100 &   0.801   (0.14)  &   10.50   (6.19)  &   0.683   (0.29)  &   0.919   (0.06)  \\
&   300 &   0.804   (0.16)  &   10.44   (6.03)  &   0.634   (0.32)  &   0.973   (0.02)  \\
&   500 &   0.813   (0.15)  &   11.47   (6.35)  &   0.644   (0.31)  &   0.982   (0.01)  \\
&   2000&   0.765   (0.14)  &   11.52   (6.12)  &   0.535   (0.29)  &   0.995   (0.01)  \\\hline
\multirow{4}*{RFE}  &   100 &   0.643   (0.16)  &   49.38   (34.36)     &   0.768   (0.34)  &   0.518   (0.35)  \\
&   300 &   0.621   (0.20)  &   188.36  (96.48)     &   0.868   (0.22)  &   0.375   (0.33)  \\
&   500 &   0.625   (0.21)  &   279.77  (152.97)    &   0.808   (0.27)  &   0.442   (0.31)  \\
&   2000&   --  &   --  &   --  &   --  \\\hline
\multirow{4}*{FRS}  &   100 &   0.640   (0.16)  &   56.76   (34.93)     &   0.836   (0.26)  &   0.544   (0.36)  \\
&   300 &   0.804   (0.14)  &   63.14   (30.11)     &   0.811   (0.29)  &   0.798   (0.10)  \\
&   500 &   0.859   (0.12)  &   69.07   (27.76)     &   0.850   (0.25)  &   0.868   (0.06)  \\
&   2000&   0.919   (0.11)  &   69.81   (28.30)     &   0.871   (0.22)  &   0.967   (0.01)  \\\hline
\multirow{4}*{Lasso}&   100 &   0.637   (0.15)  &   48.11   (23.45)     &   0.745   (0.21)  &   0.530   (0.24)  \\
&   300 &   0.669   (0.16)  &   63.98   (25.04)     &   0.548   (0.27)  &   0.791   (0.09)  \\
&   500 &   0.652   (0.17)  &   81.36   (17.67)     &   0.464   (0.30)  &   0.840   (0.04)  \\
&   2000&   0.643   (0.17)  &   90.10   (10.17)     &   0.330   (0.34)  &   0.956   (0.01)  \\\hline
\multirow{4}*{FBED} &   100 &   0.604   (0.11)  &   3.11    (1.16)  &   0.230   (0.22)  &   0.977   (0.01)  \\
&   300 &   0.584   (0.10)  &   5.34    (1.26)  &   0.184   (0.20)  &   0.984   (0.01)  \\
&   500 &   0.593   (0.11)  &   6.75    (1.39)  &   0.199   (0.22)  &   0.988   (0.01)      \\
&   2000&   0.571   (0.10)  &   11.32   (1.50)  &   0.148   (0.19)  &   0.995   (0.01)      \\\hline
\end{tabular}}
\end{table}

\begin{table}[!htp]
\centering
\small
\setlength\tabcolsep{21pt}
\caption{Performance comparison under Example 5.}
\label{tableE5}
\scalebox{0.85}{
\begin{tabular}{llcccccccc}
\hline\hline
Method  &   $p$ &   Balanced Accuracy       &   Model Size      &   TPR     &   TNR     \\\hline
\multirow{4}*{FK-RFE}   &   100 &   0.977   (0.02)  &   6.43    (3.27)  &   1.000   (0.00)  &   0.955   (0.03)  \\
&   300 &   0.991   (0.02)  &   6.36    (3.04)  &   0.998   (0.04)  &   0.985   (0.01)  \\
&   500 &   0.994   (0.02)  &   6.44    (3.05)  &   0.998   (0.04)  &   0.991   (0.01)  \\
&   2000&   0.998   (0.02)  &   6.56    (3.33)  &   0.998   (0.04)  &   0.998   (0.01)  \\\hline
\multirow{4}*{RFE}  &   100 &   0.945   (0.05)  &   11.13   (6.01)  &   0.983   (0.09)  &   0.906   (0.06)  \\
&   300 &   0.982   (0.01)  &   12.80   (7.52)  &   1.000   (0.00)  &   0.964   (0.03)  \\
&   500 &   0.978   (0.05)  &   15.55   (8.41)  &   0.983   (0.09)  &   0.973   (0.02)  \\
&   2000&   --  &   --  &   --  &   --  \\\hline
\multirow{4}*{FRS}  &   100 &   0.745   (0.18)  &   42.00   (34.81)     &   0.900   (0.20)  &   0.590   (0.35)  \\
&   300 &   0.824   (0.13)  &   61.63   (36.33)     &   0.850   (0.27)  &   0.799   (0.12)  \\
&   500 &   0.841   (0.14)  &   52.03   (36.94)     &   0.783   (0.31)  &   0.899   (0.07)  \\
&   2000&   0.953   (0.08)  &   57.77   (35.96)     &   0.933   (0.17)  &   0.972   (0.02)  \\\hline
\multirow{4}*{Lasso}    &   100 &   0.585   (0.15)  &   82.38   (28.60)     &   0.990   (0.07)  &   0.180   (0.29)  \\
&   300 &   0.563   (0.11)  &   22.50   (39.55)     &   0.200   (0.30)  &   0.926   (0.13)  \\
&   500 &   0.724   (0.15)  &   71.62   (39.04)     &   0.590   (0.30)  &   0.859   (0.08)  \\
&   2000&   0.592   (0.13)  &   33.84   (46.46)     &   0.200   (0.29)  &   0.983   (0.02)  \\\hline
\multirow{4}*{FBED} &   100 &   0.747   (0.08)  &   3.20    (1.16)  &   0.517   (0.16)  &   0.978   (0.01)  \\
&   300 &   0.743   (0.11)  &   5.43    (1.63)  &   0.500   (0.23)  &   0.985   (0.01)  \\
&   500 &   0.753   (0.10)  &   6.60    (1.67)  &   0.517   (0.21)  &   0.989   (0.01)  \\
&   2000&   0.665   (0.15)  &   8.23    (1.96)  &   0.333   (0.30)  &   0.996   (0.01)  \\\hline
\end{tabular}}
\end{table}

\section{Application}
In this section, we demonstrate the performance of the FK-RFE on a Tecator dataset gathered using the Near Infrared Transmission (NIT) principle by the Tecator Infratec Food and Feed Analyzer operating in the wavelength range of 850–1050 nm. The data consists of 240 samples and 100 predictors of absorbance channel spectra and the response of the proportion of fat in meat that is finely chopped. This data collection can be accessed at \url{http://lib.stat.cmu.edu/datasets/tecator}. In application, the proposed method is compared with the RFE, FRS, and Lasso. The FBED cannot be used here for we need to predict the response, and it is valid for feature selection only.

We randomly select 200 samples as the training set and the remaining 40 samples as the testing set. Also, in addition to the 100 predictors in the original dataset, we add 900 independent noise variables following the standard normal distribution.
We discuss the effectiveness of methods in terms of model selection performance, fitting performance, and prediction performance. Model size and wrong selection (the number of selections from generated noises) are utilized to measure model selection performance and results are shown in Table \ref{tableapmodel}. We also calculate Mean square error,
\begin{align*}
\text{MSE}=\frac{1}{n}\sum_{i=1}^{n}(\hat y_i-y_i)^2,
\end{align*}
Mean absolute error,
\begin{align*}
\text{MAE}=\frac{1}{n}\sum_{i=1}^{n}|\hat y_i-y_i|,
\end{align*}
and Mean absolute percentage error,
\begin{align*}
\text{MAPE}=\frac{1}{n}\sum_{i=1}^{n} \dfrac{|\hat y_i-y_i|}{y_i} \times 100\%,
\end{align*}
on the training and testing set, respectively. The performance of the RFE and the FRS is based on the random forest, and the results of all approaches on these three metrics are shown in Tables \ref{tableapfit}-\ref{tableappre}.

In the empirical data analysis, FK-RFE always performs better than the other methods. As one can see in Table \ref{tableapmodel}, both the model size and the wrong selection of the FK-RFE are much smaller than others. Specifically, FK-RFE reduces the wrong selection of RFE by 81\%, that of FRS by 92\%, and that of Lasso by 99\%.
In the meantime, FK-RFE also achieves the lowest MSE, MAE, and MAPE from both the training set and testing set, as shown in Table \ref{tableapfit} and Table \ref{tableappre}. For example, FK-RFE reduces the predicted MAPE of RFE by 4\%, that of FRS by 13\%, and that of Lasso by 54\%.

\begin{table}[!htp]
\centering
\small
\setlength\tabcolsep{18pt}
\caption{The performance of model selection}
\label{tableapmodel}
\scalebox{1}{
\begin{tabular}{lcccccccc}
\hline\hline
&   FK-RFE  &   RFE &   FRS &   Lasso   \\
\hline
Model Size  &   20.14   (15.80)     &   21.76   (10.47)     &   93.41   (6.19)  &   163.27  (7.84)\\
Wrong Selection &   0.15    (0.45)  &   0.80    (1.03)  &   1.90    (1.86)  &   162.86  (7.66) \\
\hline
\end{tabular}}
\end{table}

\begin{table}[!htp]
\centering
\small
\setlength\tabcolsep{19.8pt}
\caption{The performance of fitting error.}
\label{tableapfit}
\scalebox{1}{
\begin{tabular}{lcccccccc}
\hline\hline
&   FK-RFE  &   RFE &   FRS &   Lasso   \\
\hline
MSE &   0.049   (0.004)     &   0.049   (0.004)     &   0.058   (0.005)     &   0.081   (0.017)     \\
MAE &   0.151   (0.006)     &   0.154   (0.008)     &   0.173   (0.009)     &   0.225   (0.024)     \\
MAPE(\%)    &   48.60   (6.748)     &   49.32   (7.433)     &   51.80   (5.873)     &   52.94   (8.150)     \\\hline
\end{tabular}}
\end{table}

\begin{table}[!htp]
\centering
\small
\setlength\tabcolsep{17pt}
\caption{The performance of prediction error.}
\label{tableappre}
\scalebox{1}{
\begin{tabular}{lcccccccc}
\hline\hline
&   FK-RFE  &   RFE &   FRS &   Lasso       \\
\hline
MSE &   0.255   (0.103)     &   0.283   (0.095)     &   0.342   (0.116)     &   1.672   (0.306)     \\
MAE &   0.354   (0.067)     &   0.380   (0.065)     &   0.432   (0.069)     &   1.026   (0.100)     \\
MAPE(\%)    &   108.18  (54.5)  &   113.62  (55.031)    &   125.65  (52.64)     &   233.77  (117.301)   \\\hline
\end{tabular}}
\end{table}

\section{Summary}
In this paper, we propose a novel model-free two phases feature selection procedure, named FK-RFE, which is a fast and efficient technique for ultra-high dimensional complex datasets. This approach is designed to remove model limitations, reduce computational complexity and search time to extract feature subsets with the minimum number of features and obtain high accuracy. We prove that the proposed method achieves the above characteristics by the fused Kolmogorov filter and the random forest based RFE. We prove the selection consistency and $L_2$ consistency of the proposed method with regular conditions. The findings obtained from comparing with other methods in simulations and application demonstrated the superior efficiency of this procedure. As a result, this technique is useful and efficient for resolving feature selection issues.

For future work, it would be worth studying the new algorithm with more efficient filter techniques for various specific types of data, such as heterogeneous data, categorical data, data with latent variables, etc. Further, to our knowledge, the asymptotic theoretical guarantees for many machine learning techniques are still missing, and it is worth filling this gap by establishing the theories for these efficient methods.

\section*{Acknowledgement}
This work was supported by the National Natural Science Foundation of China (Grant No. 12001557); the Youth Talent Development Support Program (QYP202104), the Emerging Interdisciplinary Project, and the Disciplinary Funds in Central University in Finance and Economics.

\section*{Compliance with ethical standards}
The authors declare no potential conflict of interest. The authors declare no research involving human participants and/or animals. The authors declare informed consent.

\section*{Appendix}

\begin{proof}[Proof of Theorem 1]
As shown in Algorithm 1, let $V_0, V_1,\dots, V_{d_n-1}$ be the sequence of active sets selected during iterations, in which $V_0$ is selected by the filter phase and $V_1,\dots, V_{d_n-1}$ are obtained by eliminating one variable at each step in the wrapper phase. By the nature of the sequential procedure, this is a nested sequence, i.e.,
\begin{align*}
V_0\supset V_1\supset V_2 \supset \dots \supset V_{d_n-1}.
\end{align*}
We aim to prove the convergence of the algorithm. It suffices to show that there exists a number of steps $k\in\{0,1,2\dots,d_n-1\}$ such that the mean square error of random forest estimation under the model $V_k$ is the minimum.

As mentioned in Theorem 1 of \cite{mai2015fused}, under conditions C1 and C2, we have $\{S\subset V_0\}$. Under the assumption of the importance measure,
\[ \lim_{M\to \infty}E(\hat I(X_j))=I(X_j), \]
we have the empirical permutation importance measure is unbiased. Based on the definition of the permutation importance measure, for $j\in S^c$,
\begin{align*}
I(X_j)&=E[(Y-f(X_{(j)}))^2]-E[(Y-f(X))^2]\\
&=E[(Y-f(X))^2]-E[(Y-f(X))^2]=0.
\end{align*}
On the other hand, for $j\in S$,
\begin{align*}
I(X_j)&=E[(Y-f(X_{(j)}))^2]-E[(Y-f(X))^2]\\
&=E[((Y-f(X))+(f(X)-f(X_{(j)})))^2]-E[(Y-f(X))^2]\\
&=E[(f(X)-f(X_{(j)}))^2]+E[\e(f(X)-f(X_{(j)}))]\\
&=E[(f(X)-f(X_{(j)}))^2].
\end{align*}
The last equality follows from the assumption that $\e$ is independent of $X$ and $X_{(j)}$. This leads to $E[\e(f(X)]=0$ and $E[\e(f(X_{(j)})]=0$.
Thus, we can obtain that
\begin{equation}\label{eq importance}
I(X_j)=0 ~\text{for}~ j\in S^c ~\text{and}~ I(X_j)>0 ~\text{for}~ j\in S.
\end{equation}
Noted that at each iteration, the wrapper phase eliminates the least important variable with the smallest value of the permutation importance measure. Based on the above result \eqref{eq importance}, we have that the unimportant variables would be eliminated first. Thus, set $k=d_n-q$. We have that there exists an active subset that $V_k = S$. Under the requirement of the random forest estimator, we have
\begin{align*}
\lim_{n\to \infty}E[\bar f-f]^2=0.
\end{align*}
It means that the random forest under model $V_k$ has the best performance and thus $V_k$ can be selected as the optimal model according to the criterion, i.e., $\hat S=V_k$, completing the proof.
\end{proof}

\bibliographystyle{apalike}
\bibliography{C:/reference/reference1}
\end{document}